# A Personal Data Value at Risk (Pd-VaR) approach

Author: Luis Enríquez[1]

(Université de Lille – Universidad Andina Simón Bolívar)

November 5, 2024

## Abstract

What if the main data protection vulnerability is risk management? Data Protection merges three disciplines: data protection law, information security, and risk management. Nonetheless, very little research has been made in the field of data protection risk management, where subjectivity and superficiality are the dominant state of the art. Since the GDPR tells you what to do, but not how to do it, the solution for approaching GDPR compliance is still a gray zone, where the trend is using the rule of thumb. Considering that the most important goal of risk management is to reduce uncertainty in order to take informed decisions, risk management for the protection of the rights and freedoms of the data subjects cannot be disconnected from the impact materialization that data controllers and processors need to assess. This paper proposes a quantitative approach to data protection risk-based compliance from a data controller's and processor's perspective, with the aim of proposing a mindset change, where data protection impact assessments can be improved by using data protection analytics, quantitative risk analysis, and calibrating experts' opinions.

---

[1] Professor at Université de Lille, and Universidad Andina Simón Bolívar. Contact: luis.enriquez@univ-lille.fr, luis.enriquez@uasb.edu.ec.

# Summary

# 1. Data protection risk-based compliance

Data protection risk management can be seen as the convergence of three areas of study: data protection law, information security, and risk management. Among them, risk management seems to be the most underrated one, as personal data protection has not been considered an autonomous risk management discipline. Nonetheless, data protection risk management is a very particular area where legal risks, operational risks, and financial risks collide. Firstly, information security best practice standards have been adapted in the privacy domain, inheriting some convenient practices as they consist of project implementation guides and control risk taxonomies[2]. Yet, they have several drawbacks: they don't provide input data, they don't provide meaningful metrics, they don't provide data protection risk models, and they are very weak in legal risk assessment.

Secondly, the traditional Privacy Impact Assessments have transmitted their superficiality to Data Protection Impact Assessments, as they continue acting as checklists, totally disconnected from the main principles of risk management as an applied-science discipline. As Shapiro argued, Privacy Impact Assessments have two main problems, *"they tend to emphasize description over analysis"*[3], and *"risks are typically construed narrowly"*[4]. The result is a very immature state of the art of data protection compliance, and what is worse, the illusion of an enhanced protection of the rights and freedoms of natural persons, justified by superficial and weak data protection risk management methods. Hubbard divides risk analysts in four categories: the actuaries, the war quants, the economists, and the Management consultants[5]. Cybersecurity risk management has been mainly practiced by management consultants that unfortunately *"are also the most removed from the science of risk management and may have done far more harm than good"*[6]. The privacy and data protection world has inherited this soft approach to risk management, even though that the goal of risk management is even higher, to protect the rights and freedoms of the data subjects. Thus, it is necessary *decomposing the nature of a data protection risk-based approach (1.1), estimating the impact of a data breach on the data subjects (1.2), the riskification of Data Protection Authorities (1.3),* and, *implementing legal analytics for data protection risk management (1.4).*

---

[2]  See, INTERNATIONAL ORGANIZATION FOR STANDARDIZATION, ISO/IEC 27701:2019, ISO, 2019, and, NATIONAL INSTITUTE OF STANDARDS AND TECHNOLOGY, NIST SP 800-53 rev. 5, NIST, 2020 [online].
[3]  SHAPIRO (S.), "Time to Modernize Privacy Impact Assessment", *in Issues in Science and Technology, Vol.38, No. 1,* 2021, p.21.
[4]  *Ibid.*
[5]  HUBBARD (D.), *The Failure of Risk Management*, John Wiley & sons Inc, United States, second edition, 2020, p.104.
[6]  *Ibid.,* p.105.

**1.1. Decomposing the nature of a data protection risk-based approach.** The GDPR, and most data protection acts, follow a risk-based approach[7]. However, it is quite surprising that most of them (including the GDPR) do not even include a definition of risk. Risk may be defined as *"the potential for loss or disruption caused by an incident, and is to be expressed as a combination of the magnitude of such loss or disruption and the likelihood of occurrence of the incident"*[8]. This definition can easily be adapted to the personal data protection domain, as the materialization of a data breach will produce losses to the data subjects, and in the meantime, it will also produce losses to the data controllers and processors. Within this context, there are three data protection stakeholder groups that are deeply interconnected, the data protection authorities (regulators), the data controllers and processors (regulatees), and the data subjects (natural persons). The regulatory nature of this proactive legal approach is better understood as a meta-regulation, defined as *"the regulation of self-regulation"*[9], where data protection authorities shall control the risk management methods applied by the regulatees, with the aim of protecting the rights and freedoms of the data subjects. This meta-regulatory environment can easily fail if data protection authorities do not control data protection risk management properly, turning them into a vulnerability to the rights and freedoms of the data subjects[10].

The nature of a risk-based approach relies on probabilistic methods, where achieving 100% protection is unreal. Therefore, the main objective of risk-based compliance is to apply an effective risk management stack[11] in order to reduce data protection uncertainty. Personal data security depends on information security as a primary dependency, since information security risks are also risks to personal data. Meanwhile, a data breach will produce harm that can be represented as the financial losses due to the materialization of the risk. From a regulator's perspective, a data breach may become the proof of the data controller's lack of GDPR compliance, even though that residual risk is unavoidable[12]. From a data controller's perspective, a data breach produces primary losses

---

7   *"Risk management is at the heart of the accountability principle and of the risk-based approach"*. GUELLERT (R.), *The Risk Based Approach to Data Protection*, Oxford University Press, United Kingdom, 2020, p.152.
8   NIS 2 Directive, article 6 (9).
9   PARKER (C.), *The Open Corporation*, Cambridge University Press, Australia, 2002, p.245.
10  As Sparrow noted concerning regulatory agencies, *"they may have to invest in the construction and operation of systems designed to make the invisible visible – to show them what they otherwise would not have known"*. SPARROW (M.), *The Regulatory Craft: controlling risks, solving problems, and managing compliance*, United States, Brookings Press, 2000, p.265.
11  Freund and Jones proposed a risk management stack composed by accurate models, meaningful measurements, effective comparisons, well-informed decisions and effective risk management. See, FREUND (J.), JONES(J.), *Measuring and Managing Information Risk: a FAIR approach*, Elsevier Inc, United States, 2015, p.279.
12  "It is important to note that – even with the adoption of a risk-based approach – there is no question of the rights of individuals being weakened in respect of their personal data. Those rights must be just as strong even if the processing in question is relatively 'low risk'". ARTICLE 29 DATA PROTECTION WORKING PARTY, *Statement on the role of a risk-based approach in data protection legal frameworks Adopted on 30 May 2014*, Brussels, 2014, such as productivity, response, and replacement, and secondary losses such as competitive advantage, judgments and fines, and reputational losses[13]. Within this classification, an administrative fine would be considered as a secondary loss. From a data subject's perspective, they may suffer damages to their rights and freedoms as a consequence of a data breach, and such damages may materialize in a quantifiable impact, such as losing a job, higher fees for insurance, or any other. The individual impact shall finally be quantified by judges, with the purpose of getting compensation[14]. Yet, administrative authorities shall also consider the impact of a data protection violation on the rights and freedoms of the data subjects[15].

**1.2. Estimating the impact of a data breach on the data subjects.** Data controllers are obligated to consider the data subject's impact of a data breach within risk management, but in practice, this task is very challenging for two circumstances: different groups of vulnerable data subjects, and different data subjects' privacy values. Firstly, the only special vulnerable group of individuals established in the GDPR are children[16]. Malgieri observed that there are two moments in which vulnerability can manifest itself: *"(i) vulnerability during the data processing and (ii) vulnerability as a consequence of the data processing"[17]*. Thus, a data subjects' vulnerability may be revealed as a consequence of a data breach, and the main challenge of data protection risk management is obtaining and calibrating the input values of the probability of occurrence and the magnitude/impact of a data breach on the data subjects. My hypothesis is that data controllers don't have the competence and have not been trained in legal decision-making. Consequently, their estimations of the impact of a data breach on the data subjects may be highly disconnected from the reality. Secondly, data subjects evaluate their own privacy differently, making it very hard to get an accurate estimation of a global population from uninformed score-based estimations. Those are the reasons why a data subject's risk materialization perspective shall be included as a component of a data controller's perspective on risk-based GDPR compliance. But perhaps a better compliance strategy is applying case-based legal reasoning through data protection analytics, instead of guessing the probability of occurrence and the impact of a personal data risk.

**1.3. The riskification of Data Protection Authorities.** Estimating the impact on the rights and freedoms of the natural persons is the duty of the data protection authorities. A rationale-based

---

    p.2.
13  See, *Ibid.*, pp.66–73.
14  GDPR, article 82.
15  GDPR, article 83 (2a).
16  GDPR, article 8.
17  MALGIERI (G.), *Vulnerability and Data Protection Law,* United Kingdom, Oxford University Press, 2023, p.80.

approach to data controllers' risk-based compliance, may be to analyze and understand the controlling and sanctioning psychology of data protection authorities. Lawlor proposed decades ago that *"any system of successful prediction that is to be effective must involve not only a study of earlier decisions, but also a study of the judges who rendered them"*[18]. In the light of data protection risk management, the role of data protection authorities equals that of legitimate data protection experts that measure the impact on the rights and freedoms of natural persons, at least in a reactive manner. The quantitative study of law has had many decades of research, since the definition of jurimetrics by Loevinger[19]. Jurimetrics provide very valuable information that can be used by data controllers for their risk management compliance obligations, providing meaningful data inputs for the task of building data protection risk models. The analysis of historical data is a main input for cases of epistemic uncertainty[20] in other areas such as assurance, finance, and econometrics. Thus, using historical data as input for data protection risk assessments helps to build a general reference for compliance risk scenarios. Yet, the development of jurimetrics and legal analytics has been mainly linked to academic research and not necessarily to the risk management industry.

Furthermore, the fact that data controllers and processors are obligated to protect the rights and freedoms of natural persons does not mean that data protection authorities are disconnected from a compulsory risk transformation. If data protection law relies on risk management, data protection authorities need to get into a riskification[21] process with the aim of promoting effective risk-based compliance mechanisms. Until now, some of them have promoted soft risk management methods inherited from alleged best practices standards that unfortunately, are selling a simple approach to privacy/data protection risk management that masks the complexity of the task. Therefore, a change of mindset is required, where data protection decision-making can remain as an art only if data protection risk management is rationale-based[22]. Since risk management does not work by default, they shall promote risk methodologies based on applied-science. It is compulsory to rely on quantitative methods such as probability distributions, loss exceedance curves, conformal prediction, Monte Carlo analysis, among others. The lack of data may also be at least replaced by qualitative methods that can enhance decision-making, such as the Delphi method, the Lens method, and even the use of machine learning based models that help to develop argument retrieval methods from jurisprudence. Some of them will be tackled on in the next paragraphs.

---

18  LAWLOR (R.), "What Computers Can Do: Analysis and Prediction of Judicial Decisions", *in American Bar Association Journal*, *Vol.49, No.4, ABA*, 1963, p.340.
19  See, LOEVINGER (L.), "Jurimetrics—The Next Step Forward", *in Minnesota Law Review, Vol.33, No.5*, 1949, pp.455-493, and, LOEVINGER (L.), Jurimetrics: The Methodology of Legal Inquiry, *in 28 Law and Contemporary Problems*, Duke Law, United States, 1963, pp.5-35.
20  *"Also known as reducible or systematic uncertainty, this type originates from the lack of knowledge about the system or process under study"*. MANOKHIN (V.), *Practical Guide to Applied Conformal Prediction in Python*, Packt Publishing, United Kingdom, first edition, 2023, p.92.
21  Term used by Spina. See, SPINA (A.), "A Regulatory Mariage de Figaro", *in European Journal of Risk Regulation*, *Vol.8, No.1*, Cambridge University Press, 2017, p.89.
22  Or better said, a *"data protection rationale mindset"*. See, KOOPS (B.), "The problem with European Data Protection Law", *in International Data Privacy Law, Vol.4, Issue 4*, 2014, , p.255.

**1.4. Implementing legal analytics for data protection risk management.** Applying machine learning models to train information systems in the legal domain is not new. There are several research precedents in other legal areas where authors have applied them with the aim of predicting the behavior of courts. For instance, Katz and Bommarito used them to predict the behavior of the US federal court with good levels of accuracy[23]. Aletras and Lampos published their own research on predicting the sentences of the EU Court of Human Rights[24]. Medvedeva, Vols *et al.* also did research in predicting the behavior of the EU Court of Human Rights[25]. In their research, the main component was historical data (legal precedents) in order to forecast future court decisions. A historical analysis may be informative but still does not complete the task of risk modeling strategic, political, or macroeconomic present conditions may also influence the outcome of a legal decision. Nonetheless, historical legal analysis helps to reduce uncertainty in countries with strong jurisprudential lines, and it helps to detect bias and noise[26] in the decision-making of inaccurate data protection authorities. The use of predictive analytics in the service of risk management is not widespread, except in mature risk-based disciplines such as actuarial science and econometrics. Yet, data protection predictive analytics can provide huge benefits to data protection risk assessment in order to collect and analyze data protection informative data, as the necessary input for risk modeling.

## 2. Data protection analytics / Impact

Input data can be retrieved by using automated methods for information retrieval and argument retrieval[27]. A good start point is decomposing the problem, where the object of decomposition is an administrative fine. These factors are the ones recommended by the European Data Protection

---

23  See, KATZ (D.), BOMMARITO (M.), et al., "A General Approach for Predicting the Behavior of the Supreme Court of the United States", arXiv:1612.03473 [physics.soc-ph], 2017 [online], pp.1-15.
24  See, ALETRAS (N.), LAMPOS (V.), "Predicting judicial decisions of the European Court of Human Rights: a Natural Language Processing Perspective", *in Pee J. Computer Science 2:e93*, 2016.
25  See, MEDVEDEVA (M.), VOLS (M.), et al., "Using machine learning to predict decisions of the European Court of Human Rights", *in Artificial Intelligence and Law 2*, 2019.
26  See, KAHNEMAN (D.), SIBONY (O.), *et al., Noise A Flaw in Human Judgment*, Harper Collins Publishers, New York, 2021, p.5.
27  See, GRABMAIR (M.), ASHLEY (K.), *et al.*, "Introducing LUIMA: An Experiment in Legal Conceptual Retrieval of Vaccine Injury Decisions using a UIMA Type System and Tools", *in Proceedings of the 15th international conference on artificial intelligence and law*, 2015, pp.69-72.

Board to data protection authorities as a starting point for the calculation of an administrative fine[28]. Data protection's impact can be based on: *the turnover of the undertaking (2.1.), the category of the infringement (2.2),* and, *the seriousness of the infringement (2.3).*

**2.1. The turnover of the undertaking.** Retrieving information about the turnover of the undertaking is a good departure point for data protection risk management, in order to calibrate a range. It helps to set up range limits and discarding the absurd[29]. The following dataset shows the *mean* of the turnover of the undertaking before 2023 in France, the UK, Spain, and Ireland, with a sample space composed by aleatory chosen administrative fines in an annual turnover between €10 000 000, and €100 000 000:

|  | date | year | controller | fine | france | uk | spain | ireland |
|---|---|---|---|---|---|---|---|---|
| 1 | 05-2019 | 2019 | Sergic_SAS | 400000.0 | 400000.0 | NaN | NaN | NaN |
| 2 | 11-2019 | 2019 | Futura_International | 500000.0 | 500000.0 | NaN | NaN | NaN |
| 27 | 01-2021 | 2021 | Rancom Security Limited | 1279000.0 | NaN | 1279000.0 | NaN | NaN |
| 47 | 07-2021 | 2021 | AG2R_La_Mondiale | 1750000.0 | 1750000.0 | NaN | NaN | NaN |
| 52 | 12-2021 | 2021 | NBQ_technology | 24000.0 | NaN | NaN | 24000.0 | NaN |
| 55 | 04-2022 | 2022 | Dedalus Biologie | 1500000.0 | 1500000.0 | NaN | NaN | NaN |
| 61 | 05-2023 | 2023 | Doctissimo | 380000.0 | 380000.0 | NaN | NaN | NaN |
| 71 | 10-2022 | 2022 | EasyLife ltd | 1567000.0 | NaN | 1567000.0 | NaN | NaN |
| 95 | 12-2022 | 2022 | Virtue integrated Elder Care | 100000.0 | NaN | NaN | NaN | 100000.0 |
| 96 | 12-2022 | 2022 | A&G couriers | 15000.0 | NaN | NaN | NaN | 15000.0 |
| 106 | 03-2021 | 2021 | Irish Credit Bureau | 90000.0 | NaN | NaN | NaN | 90000.0 |

```
In [80]: sanctions_evaluations.france.mean()
Out[80]: 906000.0

In [81]: sanctions_evaluations.uk.mean()
Out[81]: 1423000.0

In [82]: sanctions_evaluations.spain.mean()
Out[82]: 24000.0

In [83]: sanctions_evaluations.ireland.mean()
Out[83]: 68333.33333333333
```

**2.2. The category of the infringement.** The second factor is the category of the infringement. The GDPR only establishes a higher category up to the 4% of the turnover of the undertaking for some GDPR violations and a lower category up to the 2% for others[30]. Nevertheless, the range is still too wide. Empirical observation shows that a more accurate category of the infringement is based on specific GDPR articles, with the limitation that the data protection authorities shall only sanction based on the article related to the highest category of the infringement[31]. The following graphic shows the results of a sample space between €100 million and €1 billion, but with a better estimation due to the addition of the category of the infringement layer:

```
sanctions_evaluations.france.mean()
19340625.0

sanctions_evaluations.uk.mean()
6641975.0

sanctions_evaluations.spain.mean()
320583.3333333333

sanctions_evaluations.ireland.mean()
229809000.0
```

---
28 EUROPEAN DATA PROTECTION BOARD, *Guidelines 04/2022 on the calculation of administrative fines under the GDPR version 1.0*, European Union, 2022 [online], accessed on 28/10/2022.
29 See, JOSEY (A.), *et al.*, *Preparation for the Open FAIR Part 1 Examination study guide*, Open Fair Foundation, United Kingdom, 2014, p.58.

**2.3. The seriousness of the infringement.** The GDPR's article 83(2) includes eleven criteria, with the first criterion weighing the impact on the data subjects, and the following ten criteria as aggravating or mitigating conditions. The limitation for risk management is that these criteria have to be estimated as a whole[32], and not by weighing each criterion. This condition is not directly useful for quantitative risk analysis unless there is a specific argument that justifies it. For instance, the following graphic shows an example by comparing the administrative fine reduction due to the COVID pandemic in the UK with an average reduction of £3 283 334 million:

| | controller | fine | france | uk | spain | article | cause | cap | number_records | item | nature |
|---|---|---|---|---|---|---|---|---|---|---|---|
| 10 | 09-2020 | Marriot | 18400000 | NaN | 24000000.0 | NaN | 1 | 51f | 9019400000 | 339000000.0 | ICO_MP_Marriot | s_confidentiality |
| 12 | 10-2020 | British airways | 20000000 | NaN | 24000000.0 | NaN | 1 | 51f | 13290000000 | 429612.0 | ICO_british_airways | s_confidentiality |
| 20 | 11-2020 | Ticket_Master | 1250000 | NaN | 1500000.0 | NaN | 1 | 51f | 11500000000 | 1500000.0 | ICO_MP_Ticketmaster | s_confidentiality |

```
turnover_medium_article_1_records_atag.uk.mean()
16500000.0

16500000 - 13216666
3283334
```

---

30  GDPR, article 83 (4).
31  *Ibid.*
32  EUROPEAN DATA PROTECTION BOARD, *Guidelines 04/2022 on the calculation of administrative fines under the GDPR version 1.0, op. cit.*, p.16.

Nonetheless, the seriousness of the infringement criteria may be useful for retrieving input data by using argument retrieval techniques. There are several methods to calibrate the experts opinions, such as the Delphi method and the Lens model. The Delphi method can be used *"when expert judgment is necessary because the use of statistical methods is inappropriate"*[33]. The Delphi method uses experts in an anonymous way, with the purpose of removing bias as much as possible. The Lens model consists of *"inviting the experts, asking them to identify a list of factors, generating scenarios with values for each factor, getting the experts' evaluation for each scenario, averaging the estimates of the experts together, and performing a logistic regression analysis with the experts' estimations"*[34]. However, my own research shows functional results with linear regression models when the goal is detecting the noise in the expert's opinions. The following graphic shows a hybrid implementation of both while calibrating an input value for data protection risk management purposes:

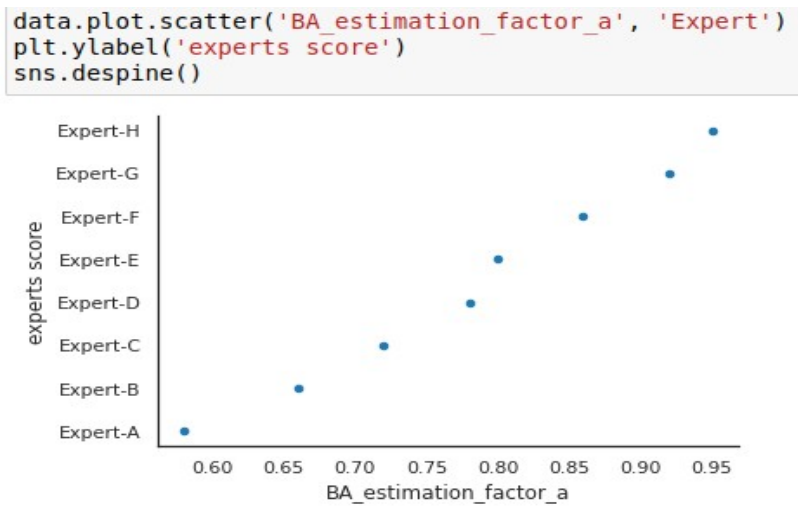

In the previous example, eight experts labeled their own opinions on the importance of a specific impact criterion from the GDPR's article 83(2A). Furthermore, understanding the sanctioning psychology of data protection authorities may also require labeling the arguments that justify each administrative fine's amount in order to train an information system. For such task, the specific criteria related to the seriousness of the infringement can be uploaded in a dataset by using Natural Language Processing:

---

[33] ROWE (G.), WRIGHT (G.), "Expert opinions in forecasting: The role of the Delphi Technique", *in* ARMSTRONG (J.) (ed.). *Principles of Forecasting*, Boston: Kluwer Academic, 2021, p.135.
[34] HUBBARD (D.), *The Failure of Risk Management*, *op. cit.*, pp.185-186.

| | legal factor | case | year | Country | administrtive fine | argument | weight |
|---|---|---|---|---|---|---|---|
| 0 | a | Marriot Hotels | 2020 | UK | 18000000 | An extremely large number of individuals were ... | 5 |
| 1 | a | British Airways | 2020 | UK | 23300000 | A significant number of individuals (429,612 ... | 5 |
| 2 | a | Karantinas | 2021 | Lituania | 12000 | the DPA found that the personal data of 677 i... | 3 |
| 3 | a | Indecemi | 2022 | Spain | 5000 | only two persons were affected by the confide... | 3 |
| 4 | a | Bank of Ireland | 2023 | Ireland | 100000 | The controller also confirmed, among other thi... | 2 |
| 5 | a | Olavs Hospital | 2021 | Norway | 67000 | A significant number of patients were affected... | 4 |
| 6 | a | Secretaria nacional para la innovacion y calidad | 2020 | Spain | 0 | The notified security breach concerned 34 affe... | 1 |
| 7 | a | Ticket Master | 2020 | UK | 1456000 | 9.4 million EEA data subjects were notified as... | 2 |
| 8 | a | Med Help | 2021 | Sweden | 1179500 | According to Computer Sweden, 2.7 million reco... | 4 |
| 9 | a | UK Cabinet Office | 2021 | UK | 582640 | It was found that the CSV file was accessed 38... | 5 |

By training and calibrating the information system with these labeling criteria, the testing phase should reflect the training criteria in the upcoming observations. However, such datasets shall be updated regularly, as decision-making circumstances may always change.

## 3. Data protection analytics / Probability of occurrence

All the exposed methods may help to obtain relevant historical data and to have an informed idea of the sanctioning psychology of each data protection authority. However, the previous graphics have been concerned about obtaining data for the impact. Historical data can also be the departure point for estimating the probability of occurrence. There are two common mistakes when estimating the probability of occurrence in the data protection area. Firstly, a probability of occurrence shall be estimated within a given time-frame[35]. It is concerning to see that several of the Data Protection Impact Assessment's software and even alleged best practices standards[36] do not include this fundamental risk-based practice. Secondly, the only way to calibrate the probability of occurrence is using applied scientific methods based on statistics, conditional probability, conformal prediction, among others. Input data may be commonly retrieved by following a *frequentist approach (3.1)*, and, *a Bayesian approach (3.2).*

**3.1. Frequentist approach.** It consists of estimating the probability of occurrence by observing the frequency of an event in a given time-frame. The only reliable way to represent it is through probability distributions[37]. Concerning GDPR risk-based compliance, we may estimate the

---

35  Freund and Jones described it as *"temporally bound probability"*. FREUND (J.), JONES (J.), *Measuring and Managing Information Risk: A FAIR Approach,* Elsevier Inc, United States, 2015, p.16.
36  Such as the ISO/IEC 29134:2017.
37  "There are different ways to represent probability distributions depending on whether they involve discrete or continuous outcomes". KOCHENDERFER (M.), WHEELER (T.), *et al., Algorithms for Decision Making,* United Kingdom, The MIT Press, 2022, p.20.

probability of occurrence of an administrative fine by the data protection authority. The following graphic shows the probability of occurrence of an administrative fine in 2023 in France in a given turnover's range between €10 million and €1 billion by using historical data and a normal probability distribution:

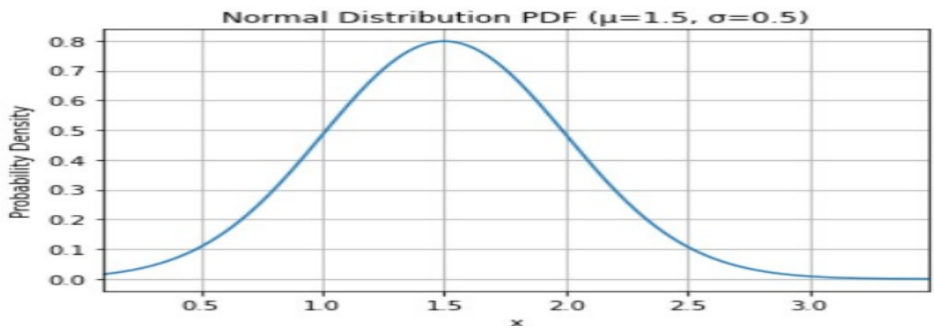

From a data controller's perspective, a frequentist approach may be used to estimate different risk scenarios, by analyzing the such as the probability of getting a data breach in a given time-frame, the probability of being controlled by the data protection authority once a data breach has happened, or the probability of receiving an administrative fine once a data controller has been controlled. The following Poisson probability distribution shows a total number of administrative fines with historical data after being controlled by the DPA in an ordinary procedure, setting the mean at 19 per year:

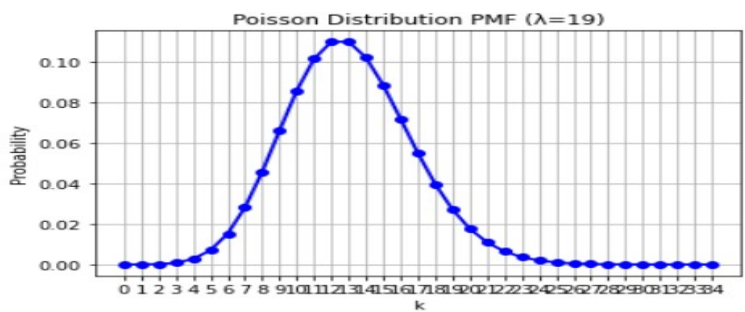

**3.2. Bayesian approach.** The Bayesian inference[38] can also be used when the probability of an event depends on another event. This is the case of the probability of getting a data breach, given that a proper Data Protection Impact Assessment has been implemented, or not. The rationales of the prior assumption may also use historical data in order to obtain meaningful outcomes:

---

[38] The advantage of Bayesian models is *"arising from scientific background, expert judgment, or previously collected data"*, and combine it *"with current data via the likelihood function to characterize the current state of knowledge using the so-called posterior distribution"*. GHOSH (S.), "Basics of Bayesian Methods", *in Methods in molecular biology*, 2010, p.153.

**db = Data breach**
**ext = External attack**
**dpia = Effective Data Protection Impact Assessment**

___________________________

**Calibrated values**

P(db | ext)=0.80
P(db | ∼ext)=0.20
P(∼db | ext)=0.20
P(ext | dpia)=0.10
P(ext|∼dpia)=0.90
P(dpia)=0.70
P(∼dpia)=0.30
P(∼ext)=0.20
P(∼db)=0.40

**Derived values**

P(ext) = P(dpia) · P(ext | dpia) + P(~dpia) · P(ext | ~dpia) = 0.34
P(db) = P(ext) · P(db | ext) + P(~ext) · P(db | ~ ext) = 0.404
P(ext | db) = P(db | ext) · P(ext) / P(db) = 0.673
P(ext | ~db) = P(~db | ext) · P(ext) / P(~db) = 0.114

**Required outcomes**

P(db | dpia) = P(ext | dpia) · P(db | ext) + P(~ext | dpia) · P(db | ~ext) = 26%
P(db | ∼dpia)=P(db | ext) · P(ext | ∼dpia) + P(db | ∼ext) · P(∼ext | ∼dpia) = 74%

Finally, given that a data breach may be caused by a breach of data confidentiality, a breach of data integrity, or a breach of data availability, a data protection officer may find it useful to use the *law of total probability*. For such a task, it is convenient to unveil the amount of administrative fines issued due to each of the three data security principles. The results in all the analyzed countries showed a predominant tendency of confidentiality data breaches, which changes the outcome of data breaches into the outcome of sanctioned data breaches:

**SOC information about information security incidents in 2023 (in a specific data controller):**

Confidentiality (C) = 76% ; P(C) = 0.76
Integrity (I) = 16.5% ; P(I) = 0.165
Availability (A) = 7.5% ; P(A) = 0.075

**Distribution (D) of administrative fines in the EU based in the data security principles (just an scenario):**

Confidentiality administrative fines = 20% ; P(D | C) = 0.2
Integrity administrative fines = 8% ; P(D | I) = 0.08
Availability administrative fines = 5% ; P(D | A) = 0.05

Probability of getting an administrative fine by a data breach = P(D) = P(C) P(D | C) + P(I) P(D | I) + P(A) P(D | A)

**Results:**

P(D) = 0.16895
P(C | D) = 89,94% of getting fined by confidentiality data breaches
P(I | D) = 7.81% of getting fined by integrity data breaches
P(A | D) = 2.22% of getting fined by availability data breaches

# 4. Conformal prediction and the jurimetrical Pd-VaR

All the exposed methods may help to obtain data for feeding the impact and likelihood metrics, by using historical data and expert calibration techniques. The next step is to represent data protection risks in an informative manner. This paper promotes the idea of using a Personal Data Value at Risk approach[39]. The Pd-VaR relies on the idea of the traditional Value at Risk (VaR)[40] consisting of three elements: Estimating the worst loss if the risk materializes in a given time-frame at a certain confidence level. The worst loss may be obtained by using the impact-based metrics already presented, and the given time-frame is necessary as the probability of occurrence may change among different periods of time. The VaR model evolved into the Cyber Value at Risk model with several proposals[41]. Yet, the importance is changing the way we communicate risk from a subjective manner into an objective and informative one[42]. Thus, this paper proposes *implementing the jurimetrical PdVaR (4.1),* and, *fixing the confidence level with conformal prediction (4.2).*

**4.1. Implementing the jurimetrical PdVaR.** The same VaR logic may be used for the privacy/data protection area in order to obtain meaningful rationales for Data Protection Impact Assessments. In such direction, instead of using superficial linear methods such as multiplying the impact and the frequency values[43], risk matrices and heat maps[44], a Pd-VaR shall express in a better way the real meaning of a risk. From a data subject's perspective, the Pd-VaR may be expressed as "If an administrative *fine (if controlled) happens next year, there is a 90% chance that the sanctioning amount will be between €300 000 and €400 000"*[45]. This inference may be the result of a particular individual estimating his own losses. However, from a data controller's perspective, it is very subjective to guess about the material impact of a data breach on particular data subjects. A better strategy is to focus on data protection as a compliance risk, where it becomes more accurate to

---

[39] *"The jurimetrical Pd-VaR shall be the prior information retrieved from the administrative fines issued by the Data Protection Authorities"*. ENRIQUEZ (L.), *Personal data Breaches: towards a deep integration between information security risks and GDPR compliance risks*, th., Université de Lille, France, 2024, [online], p.225.
[40] See, BALLOTA (L.), FUSAI (G.), "A Gentle Introduction to Value at Risk", University of London, 2017, pp.36-37.
[41] A remarkable Cyber Value at Risk initiative was presented by the World Economic Forum in 2015. See, WORLD ECONOMIC FORUM, *Partnering for Cyber Resilience Towards the Quantification of Cyber Threats*, WEF, 2015.
[42] Hubbard and Seiersen proposed this mindset change in the cybersecurity risk management domain. See, HUBBARD (D.), SEIERSEN (R.), *How to Measure Anything in Cybersecurity Risk*, John Wiley & sons Inc, United States, 2016, p.36.
[43] See, KEMP (M.), KRISCHANITZ (C.), *Actuaries and Operational Risk Management*, Actuarial Association of Europe, 2021 [online], p.31.
[44] *"A risk matrix with more than one "color" (level of risk priority) for its cells satisfies weak consistency with a quantitative risk interpretation if points in its top risk category represent higher quantitative risks than points in its bottom category"*. COX (L.), "What's Wrong with Risk Matrices", *in Risk Analysis, Vol.28, No.2,* 2008, p.501.
[45] ENRIQUEZ (L.), Personal data Breaches: towards a deep integration between information security risks and GDPR compliance risks", th., Université de Lille, France, 2024, [online], p.264.

understand the data protection authorities' sanctioning psychology instead of guessing the impact on the data subjects. Therefore, the jurimetrical Pd-VaR consists of all the data obtained by the use of data protection analytics over the analysis of existing administrative fines. The following graphics show the jurimetrical Pd-VaR of a company in France for 2023, with a previously computed historical VaR of €95 000 and €2 million:

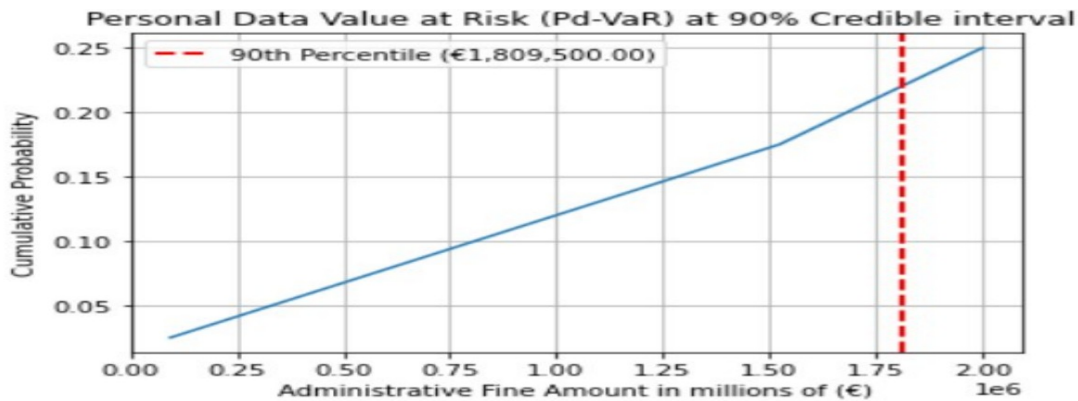

However, this historical estimation may be improved, since the 90% confidence level based on the turnover's range is logical, but fragile. As Morey and Hoekstra observed, *"any author who chooses to use confidence intervals should ensure that the intervals correspond numerically with credible intervals under some reasonable prior"*[46]. Building credible intervals can be improved in the light of predictive analytics and machine learning models, since *"the AI/ML are fundamental to move beyond the drawbacks of Cy-VaR models that mainly apply Bayesian and frequentist methods"*[47]. This is where conformal prediction becomes the best alternative, as it offers *"valid confidence measures for individual predictions"*[48]. In a nutshell, conformal prediction "is a straightforward way to generate prediction sets for any model"[49]. Implementing conformal prediction in the privacy/data protection domain is fully aligned with the idea of analyzing historical data from administrative fines and using it for risk-based compliance.

---

46 MOREY (R.), HOEKSTRA (R.), *et al.*, "The fallacy of placing confidence in confident intervals", *in Psychon Bull Rev 23*, Springer, 2016, p.118.
47 ALBINA (O.), "Cyber Risk Quantification: Investigating the Role of Cyber Value at Risk", *in Risks 9.10*, 2021, p.2.
48 MANOKHIN (V.), *Practical Guide to Applied Conformal Prediction in Python,* Packt Publishing, United Kingdom, first edition, 2023, p.27.

**4.2. Fixing the confidence level with conformal prediction.** Nevertheless, the data retrieved from administrative fines may have some limitations. Firstly, they usually have a heteroscedasticity condition, as there is a big range between the lowest limit and the highest one. Secondly, data scarcity is very common, as some data protection authorities have issued a few administrative fines. Thirdly, some administrative fines may show higher levels of bias and noise than others, as they are the result of human decision-making. Yet, conformal prediction is a convenient solution to deal with such limitations and much more efficient for regression problems than Bayesian methods, ensemble methods, and direct interval estimation methods[50]. The following example shows a very small dataset of ten administrative fines sanctioned by the CNIL in a certain turnover range and by the same category of the infringement[51]:

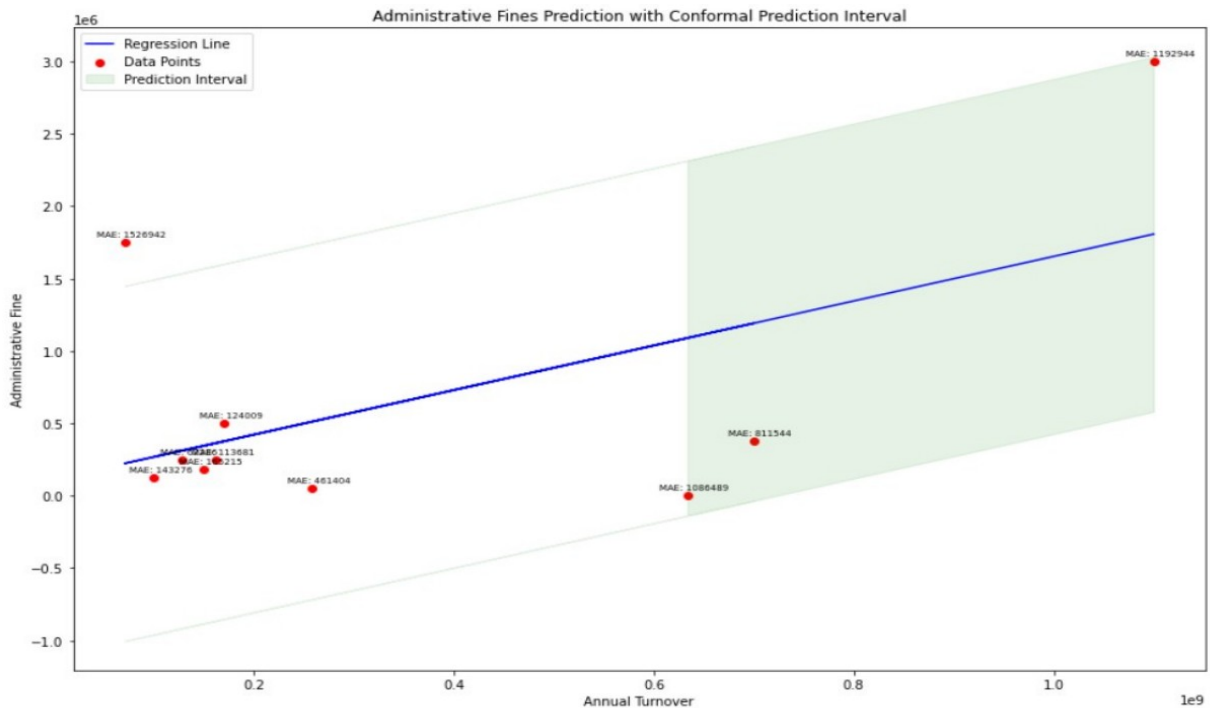

---

[49] ANGELOPOULUS (A), BATES (S.), "A Gentle Introduction to Conformal Prediction and Distribution-Free Uncertainty Quantification", arXiv:2107.07511 [cs.LG], 2022 [online], p.4.
[50] See, MANOKHIN (V.), *Practical Guide to Applied Conformal Prediction in Python, op. cit.*, p.97.
[51] Since it is a very small dataset, a method based on "transductive conformal predictors" has been applied. See, VOVK (V.), "transductive conformal predictors" *in 9th Artificial Intelligence Applications and Innovations (AIAI)*, 2013 [online].

Using transductive conformal prediction can still work for scarce data, as setting the 90 percentile confidence interval will only exclude one legal precedent. Yet, the graphic shows that there are historical precedents that may be considered far from the main data concentration. Therefore, perhaps a more credible confidence interval may be setting the confidence level at the 70th percentile. On the other hand, when a dataset is larger, the solution may be using inductive conformal prediction[52], since the dataset can be divided into quantiles and there is enough data to include a calibration set between the training and testing sets. Within this proposed methodology, a jurimetrical Pd-VaR may be obtained by using the historical analysis of administrative fines at a confidence interval determined by conformal predictors. The results are a general overview of the Personal Data Value at Risk's circumstances in a specified EU country. However, it is necessary to combine it with the specific circumstances of a particular data controller/processor. For such tasks, it is compulsory to create data privacy/protection risk models.

## 5. Customizing the FAIR model for obtaining the calibrated Pd-VaR

The calibrated Pd-VaR combines the jurimetrical Pd-VaR exposed in the previous paragraph, but combined with the current situation of a specific data controller. Modeling data protection risk relies on its own multidimensionality, which at least includes the legal risk dimension, the operational risk dimension, and the financial risk dimension. In cybersecurity risk scenarios, the FAIR model[53] has become the most popular cyber risk ontology because it suits an applied scientific risk-based approach and can merge the legal, operational, and financial risk dimensions. The FAIR model uses a Monte Carlo method[54] that is represented in a Beta Pert probability distribution[55] with three parameters: *minimum, maximum* and *most likely*. Yet, in the traditional model we may consider *Personal data administrative fines as secondary losses (5.1),* but in some circumstances may be more convenient to consider *Personal data administrative fines as primary losses (5.2.)*

**5.1. Personal data administrative fines as secondary losses.** A data breach will produce at least six types of losses: productivity, incident response, asset replacement, competitive advantage, fines and judgments, and reputational losses[56]. All these improvements are well established as primary and secondary losses in the FAIR model. The following example shows an implementation of a confidentiality data breach risk scenario where the probable administrative fine is part of the 'fines and judgments' secondary loss.

---

52 *"A set of distribution-free and model agnostic algorithms devised to predict with a user-defined confidence with coverage guarantee"*. SOUSA (M.), "Inductive Conformal Prediction: A Straightforward Introduction with Examples in Python", arXiv:2206.11810v4 [stat.ML], 2022 [online], p.1.
53 See, FREUND (J.), JONES (J.), *Measuring and Managing Information Risk: A FAIR Approach*, Elsevier Inc, United States, 2015.
54 *"The Monte Carlo method is a simple computer technique based on performing numerous fictitious experiments with random numbers"*. MENCIK (J.), "Monte Carlo Simulation Method", *in book Concise Reliability for Engineers*, University of Pardubice, IntechOpen, Czech Republic, 2016, p.127.
55 See, FREUND (J.), JONES (J.), *Measuring and Managing Information Risk: A FAIR Approach*, *op. cit.,* p.28, pp.99-101.

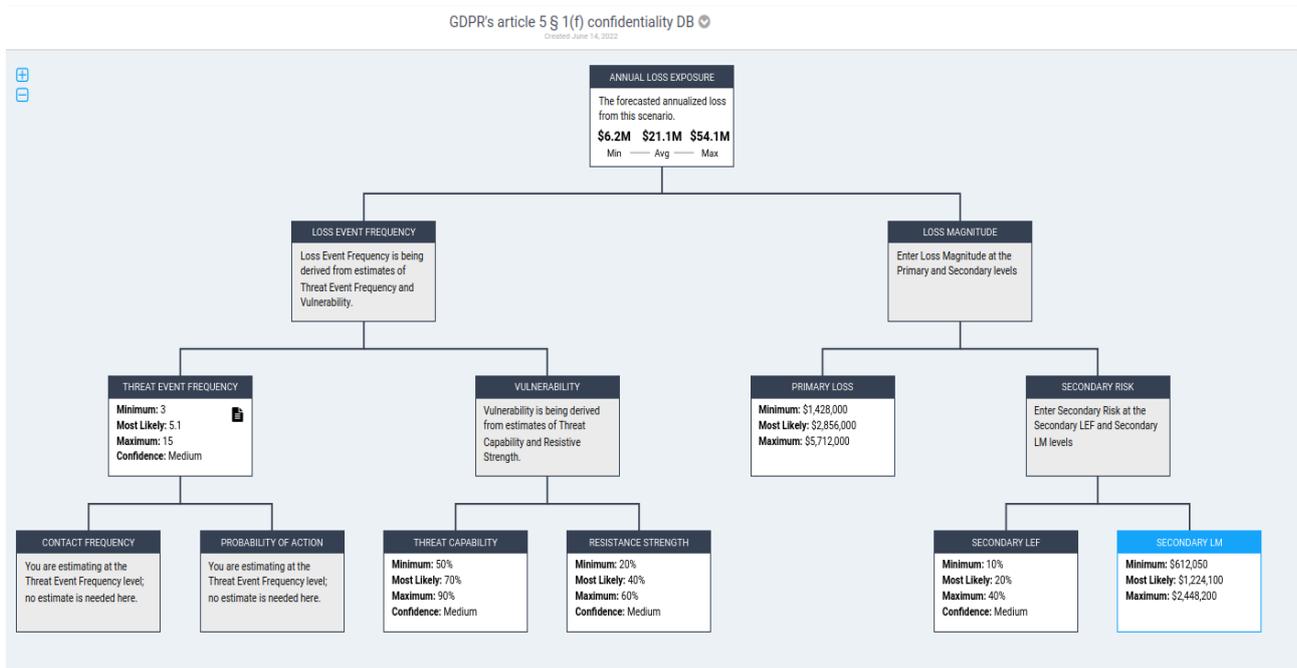
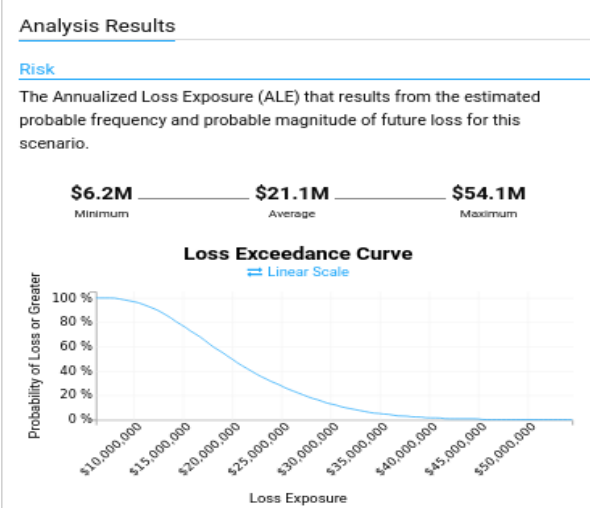

The flexibility of the FAIR model ontology, makes it a very convenient one for personal data protection risk management. Nonetheless, there are two drawbacks: the need of calibrating the input values of the maturity state of GDPR compliance within the Secondary Loss Event Frequency (SLEF), and the need of calibrating the magnitude only from the GDPR's administrative fines.

---

56   *Ibid.*, pp.66-73.

**5.2. Personal data administrative fines as primary losses.** The risk of loss due to administrative fines requires its own risk model. This can be solved with a FAIR model customization for personal data protection, where the administrative fine is the Primary Loss[57], and the Data Protection Authority is the threat community[58]. In such GDPR compliance risk scenario, the secondary losses[59] may belong to other loss types, such as reputational losses. Likewise, considering the administrative fine as the primary loss, opens the possibility of merging the jurimetrical Pd-VaR, and the calibrated Pd-VaR within the Loss Event Frequency dimension. The following graphic shows a customized version of the FAIR model:

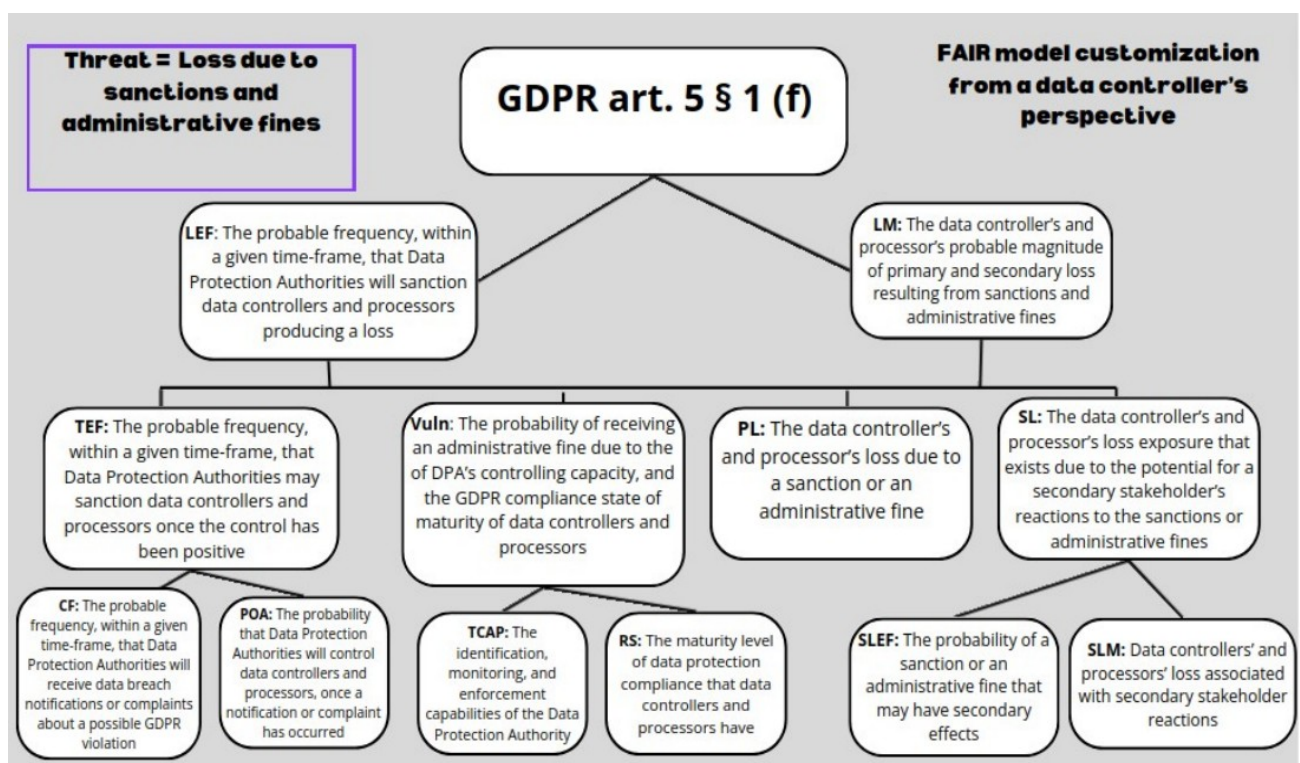

---

57 *"The data controller's and processor's loss due to a sanction or an administrative fine"*. ENRIQUEZ (L.), *Personal data Breaches: towards a deep integration between information security risks and GDPR compliance risks*, th., Université de Lille, France, 2024, [online], p.273. Also consider that an administrative fine could also consist on a temporal or definitive ban on personal data processing in some circumstances.

58 Considering the supervisory authority as a threat should not be interpreted as something negative, because their mission is to *"monitor and enforce the application of this Regulation"*. GDPR, article 57(1a).

59 *"The data controller's and processor's loss exposure that exists due to the potential for a secondary stakeholder's reactions to sanctions or administrative fines"*. ENRIQUEZ (L.), *Personal data breaches: towards a deep integration between information security risks and GDPR compliance risks*, th., Université de Lille, France, 2024, [online], p.273.

The Threat Event Frequency[60] input values are derived from the jurimetrical Pd-VaR, and then merged with the particular situation of a data controller, which comes from the Vulnerability[61] branch, derived from the Threat Capability[62], and the level of resistance strength[63] of the data controller or processor. The result is the calibrated Loss Event frequency, which will be merged with the Magnitude, in order to obtain a quantifiable level of the risk. The outcomes of this customized personal data protection model will become the input value for considering data protection administrative fines', as part of a data breach's secondary losses in the traditional FAIR model ontology.

## 6. Conclusions

This paper has presented the concept of the Personal Data Value at Risk (Pd-VaR), as the rationale of personal data protection risk management. Data protection analytics has been presented as the right approach to generate meaningful data, in order to construct data protection metrics. The jurimetrical Pd-VaR has been established as the results of the quantitative and qualitative analysis of existing administrative fines, where conformal prediction becomes the most valuable method to determine confidence intervals. The calibrated Pd-VaR is the result of merging the jurimetrical Pd-VaR, with the current situation of a data controller, especially considering the threat capacity, and the resistance strength factors. The result provides meaningful rationales not only for the 'risk' sections of a Data Protection Impact Assessment, but to all GDPR compliance obligations.

However, the role of data protection authorities is crucial, as underperforming Data Protection Authorities become a vulnerability for the rights and freedoms of the data subjects. Since data controllers and processors do not have the training or the competence to measure the impact of a data breach on the data subjects, they can still analyze how the data protection authorities are measuring them. Nevertheless, data protection authorities shall embrace a risk transformation that allows them to have better estimations of different kinds of the data subjects' impacts. Thus, data protection risk management needs to keep evolving towards a mature risk-based approach, and privacy uncertainty quantification is gradually becoming a must.

---

60 *"The probable frequency, within a given time-frame, that Data Protection Authorities will sanction data controllers producing a loss". Ibid.*, p.272.
61 *"The probability of receiving an administrative fine due to the of Data Protection Authority's controlling capacity, and the GDPR compliance state of maturity of data controllers and processors". Ibid.*
62 *"The identification, monitoring, and enforcement capabilities of the Data Protection Authority". Ibid.*
63 *"The maturity level of data protection compliance that data controllers and processors have". Ibid.*

Furthermore, the Pd-VaR could also be forecasted from a data subject's perspective. This argument relies on the actual material damage that data subjects suffer due to a data breach. The difficulty of estimating the impact on different types of data subjects may be reduced by the use of algorithm bias in order to interpret how Data Protection Authorities are approaching such data protection vulnerabilities. That is the central theme of a next paper.

LOEVINGER (L.), "Jurimetrics—The Next Step Forward", in Minnesota Law Review, Vol.33, No.5, 1949, pp. 455 493.

MALGIERI (G.), *Vulnerability and Data Protection Law*, Oxford University Press, 2023, 271 p.

MANOKHIN (V.), *Practical Guide to Applied Conformal Prediction in Python*, Packt Publishing, United Kingdom, first edition, 2023, 217 p.

MEDVEDEVA (M.), VOLS (M.), *et al.*, "Using machine learning to predict decisions of the European Court of Human Rights", *in Artificial Intelligence and Law 2*, 2019, pp. 237-266.

MENCIK (J.), "Monte Carlo Simulation Method", *in book Concise Reliability for Engineers*, University of Pardubice, IntechOpen, Czech Republic, 2016, pp. 127–136.

MOREY (R.), HOEKSTRA (R.), et al., "The fallacy of placing confidence in confident intervals", *in Psychon Bull Rev 23*, Springer, 2016, pp. 103-123.

NATIONAL INSTITUTE OF STANDARDS AND TECHNOLOGY, NIST SP 800-53 rev. 5, NIST, 2020 [online].

PARKER (C.), *The Open Corporation*, Cambridge University Press, Australia, 2002, 362 p.

Regulation (EU) 2016/679 of the European Parliament and of the Council of 27 April 2016 on the protection of natural persons with regard to the processing of personal data and on the free movement of such data, and repealing Directive 95/46/EC (General Data Protection Regulation), OJEU L 119, 27 April 2016.

SHAPIRO (S.), "Time to Modernize Privacy Impact Assessment", *in Issues in Science and Technology*, Vol.38, No.1, 2021, pp. 19-22.

SOUSA (M.), "Inductive Conformal Prediction: A Straightforward Introduction with Examples in Python", arXiv:2206.11810v4 [stat.ML], 2022 [online], pp. 1-6.

SPARROW (M.), *The Regulatory Craft: controlling risks, solving problems, and managing compliance*, Brookings Institution Press, United States, 2000, 346 p.

SPINA (A.), "A Regulatory Mariage de Figaro", *in European Journal of Risk Regulation*, Vol.8, No.1, Cambridge University Press, 2017, pp. 88-94.

VOVK (V.), "transductive conformal predictors" in 9th Artificial Intelligence Applications and Innovations (AIAI), 2013 [online], pp. 348-360.

WORLD ECONOMIC FORUM, *Partnering for Cyber Resilience Towards the Quantification of Cyber Threats*, WEF, 2015 [online].